\documentclass[journal]{IEEEtran}
\ifCLASSINFOpdf
\else
\fi

\usepackage{amsmath}
\usepackage[section]{algorithm}
\usepackage{algorithmic}
\usepackage{multirow}
\usepackage{epsfig,subfigure}
\usepackage{array}

\begin{document}
%

%
%
%
%
%
\title{Improving Aviation Safety using Synthetic Vision System integrated with Eye-tracking Devices}

\author{Mingliang Xu,Yibo Guo*, Bailin Yang, Wei Chen, Pei Lv, Liwei Fan, Bin Zhou}
\thanks{Mingliang Xu, Yibo Guo, Pei Lv and Binzhou are with the School of Information Engineering, Zhengzhou University, China. E-mail: ieybguo@zzu.edu.cn.}
\thanks{Bailin Yang is with Zhejiang Gongshang University, Hangzhou, China.}
\thanks{Wei Chen is with Zhejiang University, Hangzhou, China.}
\thanks{Liwei Fan is with Luoyang Electrooptical Equipment Research Institute, Luoyang, China.}
%
%
%

\markboth{IEEE Transactions on Systems, Man, and Cybernetics: Systems,~Vol.~XX, No.~X, 2018}%
{Shell \MakeLowercase{\textit{et al.}}: Bare Demo of IEEEtran.cls for IEEE Journals}
%



\maketitle

\begin{abstract}
. By collecting the data of eyeball movement of pilots, it is possible to monitor pilot's operation in the future flight in order to detect potential accidents. In this paper, we designed a novel SVS system that is integrated with an eye tracking device, and is able to achieve the following functions:1) A novel method that is able to learn from the eyeball movements of pilots and preload or render the terrain data in various resolutions, in order to improve the quality of terrain display by comprehending the interested regions of the pilot. 2) A warning mechanism that may detect the risky operation via analyzing the aviation information from the SVS and the eyeball movement from the eye tracking device, in order to prevent the maloperations or human factor accidents. The user study and experiments show that the proposed SVS-Eyetracking system works efficiently and is capable of avoiding potential risked caused by fatigue in the flight simulation.
\end{abstract}


%
\IEEEpeerreviewmaketitle

\section{Introduction}

Synthetic Vision Systems (SVS) are designed to provide 3D terrain images of the aircraft's future routine, in order to enhance the awareness of potential Controlled Flight into Terrain (CFIT) accidents for the pilots\cite{1}\cite{2}\cite{3}. The images are composed by a virtual environment of the outside terrains and various symbols representing the relevant information that may impact the aircraft in the flight path, such as a real-time navigation route, the buildings and water that are beyond the sight \cite{3,4}, and the depiction of traffic in the airspace ahead of the pilot \cite{5}. Through the virtual tunnel shown on the SVS displays, the pilots may detect the hidden obstacles before it is too late to react, or be aware of the potential risk of their operations that may endanger the flight.

%

In recent times, with the development of eye tracking technology that has been widely used in the flight training simulations, the original SVS techniques are also benefited via introducing the eye tracking devices into the real time aviation environment. In \cite{6} and \cite{7}, eye-tracking data were collected from pilots in order to investigate their concentration during the flight. On the other hand, by collecting the data of eyeball movement of pilots, it is possible to monitor pilot's operation in the future flight in order to detect potential accidents. In this paper, we designed a novel SVS system that is integrated with an eye tracking device, and is able to achieve the following functions:

\begin{itemize}
\item A novel method that is able to learn from the eyeball movements of pilots and preload or render the terrain data in various resolutions, in order to improve the quality of terrain display by comprehending the interested regions of the pilot.
\item A warning mechanism that may detect the risky operation via analyzing the aviation information from the SVS and the eyeball movement from the eye tracking device, in order to prevent the maloperations or human factor accidents.
\end{itemize}

The novel SVS system integrated an eye tracking device into an ordinary SVS system. A novel SVS-Eyetracking (S-E) architecture is proposed in this paper. Based on the proposed architecture, a novel algorithm of preloading and rendering the terrain data of multiple resolutions concerning the eyeball movements of pilots is illustrated. The proposed algorithm is able to transmit minimum terrain data during the data preloading period and render the polygonal tessellation considering the real-time eyeball movements. According to the questionnaires of user study, the proposed algorithm is capable of improving the driving experience of pilots.

The warning mechanism of fatigue/maloperation detection is another major feature of the proposed system. The fatigue detection via eye tracking in vehicle driving has been widely investigated by the community of smart transportation and human machine interface. However, the fatigue detection on an aircraft, especially the airline flights, is different from the detections on ordinary vehicles. The aviating operations based on the flight status are much more complicated than driving the vehicle on the road, and the maloperations are much more difficult to be recognized. In this case, we propose a novel warning mechanism based on the proposed SVS-Eyetracking architecture that is able to detect the potential risky operations on the current flight status in a non-intrusive way with good accuracy. The novel mechanism may also perceive pilots' fatigue and prevent human-factor accidents by warning them.


The paper is organized as follows: the related work of SVS systems and eye tracking devices are in Section 2; the proposed S-E architecture is illustrated in Section 3; the optimal data preload algorithm is proposed in Section 4; the new warning mechanism is illustrated in Section 5; the experiences and user study are in Section 6; the conclusion of this paper is in Section 7.

\section{Related work}

\subsection{Terrain Representation}

The measuring and representing terrain textures has been widely studied during the last decades \cite{3,11,19,27,28,29,20,21,22,23,24,25,26}. The terrain generated on a SVS display device is a continuous mesh of polygons stored in a binary tree. The general solution of terrain LOD is using multi-resolution to control the data pre-loading and rendering process. There are many related surveys and references in \cite{3,11,19,27,28,29}. The terrain data are updated in real time, and the amount of pre-loaded data is incrementally changed according to the height and velocity of the aircraft. The first applied method of dealing with massive terrain data is the restricted quadtree triangulation (RQT) algorithm. In \cite{5}, the method of triangle stripping cost models is presented. However, the earlier terrain triangulation techniques depended on greedy algorithms do not support realtime or view-dependent scenarios \cite{6,15}. In \cite{4}, the data meshes are subdivided, and the problem of irregular meshes are converted into smoother surfaces. Based on these former works, the method of wavelet analysis to terrain LOD are proposed in \cite{16}. This method supports good coherence for a movable and overlooking point of view, but if does not gurarantee the error bounds on fine-grained vertex deletion.

The recent researches on the terrain visualization are focused on dealing with the hierarchical triangulated meshes. In article \cite{11}, the method of triangle-bintree mesh is chosen. The mesh of their method is very like the same mesh in our method. The simple bintree structure is not recognized nor supported the split and merge operations. Therefore, special care must be taken to handle the continuity and coherence of maintain meshes. Meanwhile, with the requirement of better display performance, the frame rate is increased tremendously, which greatly increased the amount of pre-loaded data meshes. In \cite{14,22}, a hierarchical triangulated-irregular network (TIN) data structure with ¡°near/far¡± annotations for vertex morphing is described. These methods also include a queue-driven top-down refinement procedure for building the triangle mesh for a frame. However, The first version of TIN does not consider the automatic morphing procedures, and the memory requirements are still high for each multi-resolution meshes. The better version of a quadtree sturcture is proposed in \cite{4}, which preprocesses the height of the meshes in unified grids. The pre-processing phase computes the vertices at each quadree level, and the vertices are fitted to an approximate least-squares of the below level. The method also applies a priority queue in order to refine the quadtree from top levels to the root levels. The disadvantage of the proposed method is that it does not consider frame-to-frame coherence, while only one type of error metric is applied in the structure.

The recent development of fine-grained LOD representation techniques can handle the ordinary TIN data meshes\cite{9,10,12,13}. The new methods are able to preprocess the progressive mesh representation based on view dependent refinement, and the algorithms are allowed error metrics which are designed for flexible point of view. Moreover, detail reduction based on nested Gauss-map normal bounds are shown in their works. In \cite{16}, thin triangles are removed so that the error rate can be significantly reduced. New metrics for avoiding edge-collapse operations are proposed in \cite{15}, however, the heuristic estimation cannot justify the geometric distortions well. In \cite{17,19}, a novel feedback technique is proposed in order to process the rough rate regulations. However, the frame-to-frame consistency are not fully considered and the time complexity and cost are still depending on the size of meshes.

\begin{figure*}[htb]
  \centering
  \includegraphics[width=\linewidth]{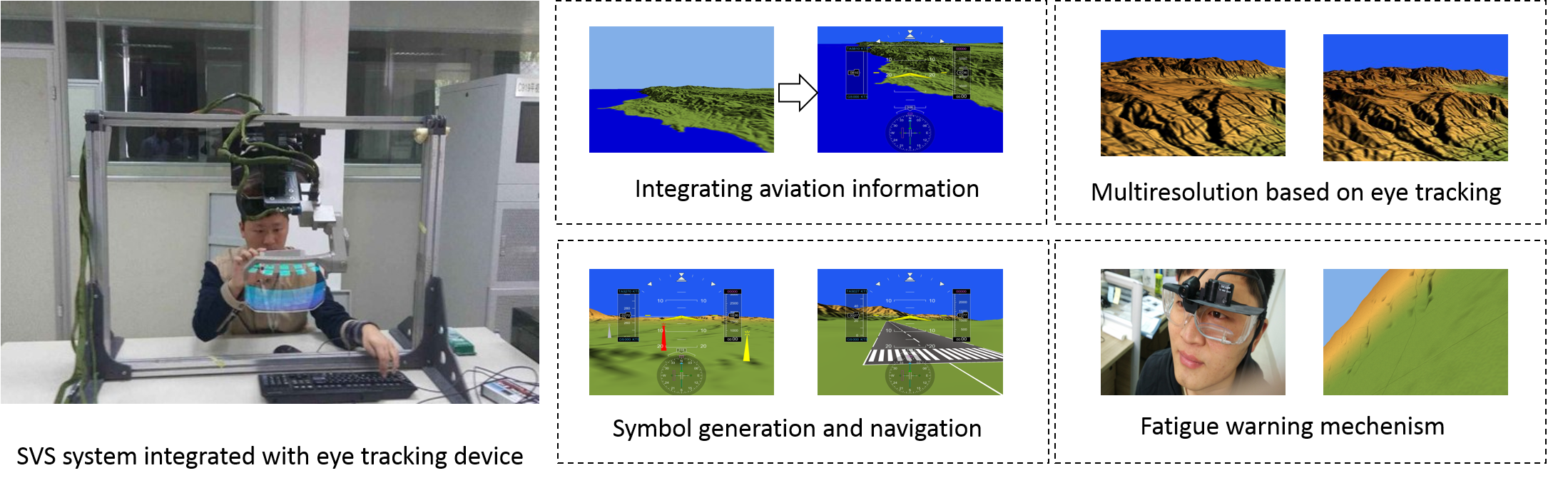}
  \caption{Synthetic Vision Systems Integrated with eye tracking device. It preserves the original SVS functions and supports customized multi-resolution based on eye tracking as well as fatigue detection mechanism.}
	\label{fig:teaser}
\end{figure*}

\subsection{Fatigue Detection}

Fatigue detection via tracking the driver's eye has been proved to be an efficient method\cite{18}. The expressions on the face such as eyelid motions, yawning, staring, or gaze are visual cues of fatigue. Eye tracking devices are installed in order to accumulate the eyeball movement information of the pilots. According to researches \cite{21,23}, most of the fatigues-realated information can be obtained from the driver's eyeball movements. There are already plenty of metrics of fatigues related to eyeball movement information being developed.  Many commercial fatigue detection systems have been developed for vehicles, such as AntiSleep developed by Smart Eye AB and Driver State Sensor (DSS) developed by Seeing Machines\cite{23}. By analyzing the timely records of the eyeball movements in a short time period, as well as the driving information collected from the SVS system, the fatigue status of the pilots can be easily justified.
 


\section{System Overview}

The integrated SVS-Eyetracking system architecture contains two modules: the SVS module, including the aviation status analyzer, the terrain data server, the display devices, and etc; the eye tracking module, including the eyeball movement tracking device, the human status analyzer, and the warning device. The architecture of the system is shown in fig 1.

The SVS module of the system contains the components of an ordinary SVS system. According to NASA SVS concept(2009), the ordinary SVS systems are composed of the sensors and database servers, the embedded computation server, and the display devices. The sensors and database servers include the weather radar, the radar altimeter, forward looking infrared sensor, and on-board synthetic vision databases. The embedded computation server analyzes the status of the aircraft in real time. It is also responsible for the image object detection and fusion, image enhancement, and terrain data rendering. The display devices are head-mounted or screen-based. In our proposed architecture, several new components are added to the embedded computation servers and the terrain data servers. New embedded hardware chips and software interfaces are installed in the ordinary SVS system in order to communicated with the eye tracking module.



The eye tracking device records three kinds of eyeball movements of the pilots: fixations, saccades, and pursuit movements. It provides the trail of the gazing spots on the terrain image to the SVS computation server, and by analyzing it with the flight status information from the SVS module, predicts the high resolution area of the terrain on the screen. On the other hand, by comparing the eyeball movements of the previous records with the present ones, the warning device may find out whether pilots are in fatigue, and remind them when the SVS module indicates the flight is in abnormal status.

\begin{figure}[htb]
 \centering
  \includegraphics[width=1.0\columnwidth]{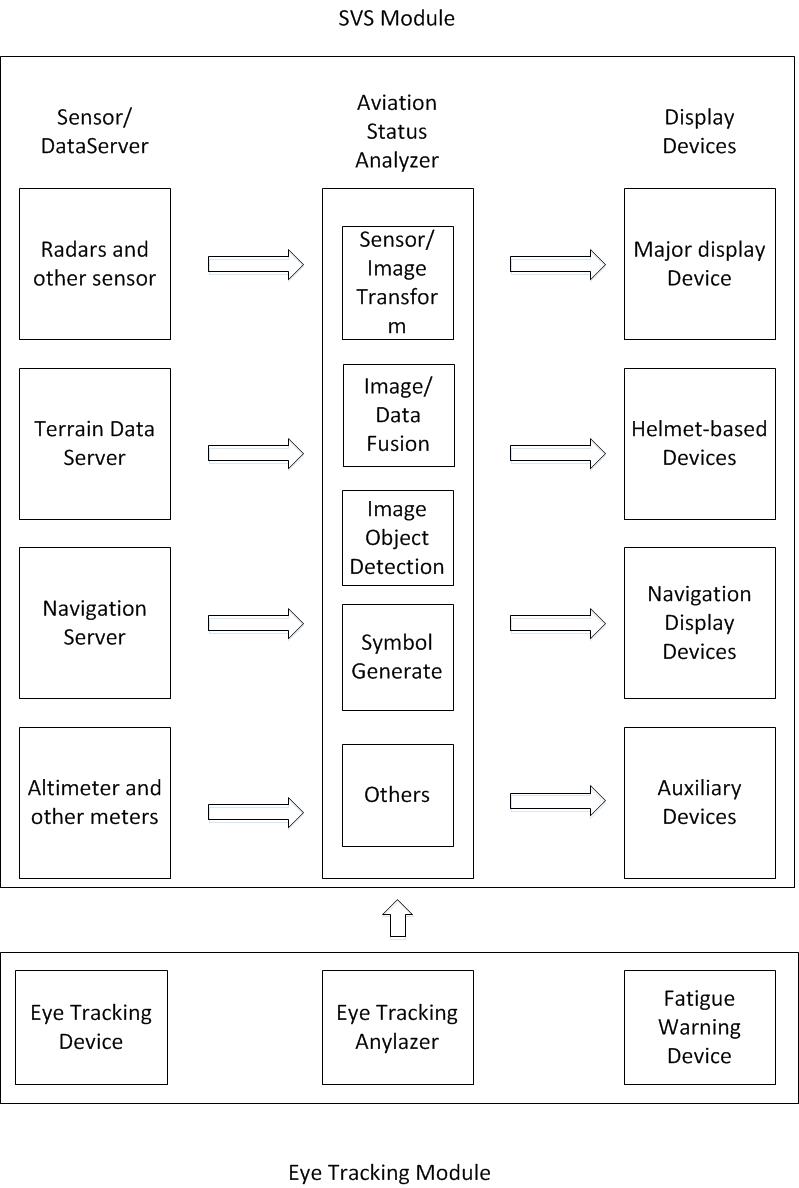}
\caption{The SVS-Eyetracking Architecture}\centering
  \label{fig:e3dg}
\end{figure}

\section{Data Preloading and rendering Algorithms}


In this section, we present a data preload algorithm that is designed to preload and render minimum terrain data from terrain data server to the computation server according to the eye-tracking information of the pilot. The procedure of the whole system is shown in figure 2.

\begin{figure}[htb]
 \centering
  \includegraphics[width=1.0\columnwidth]{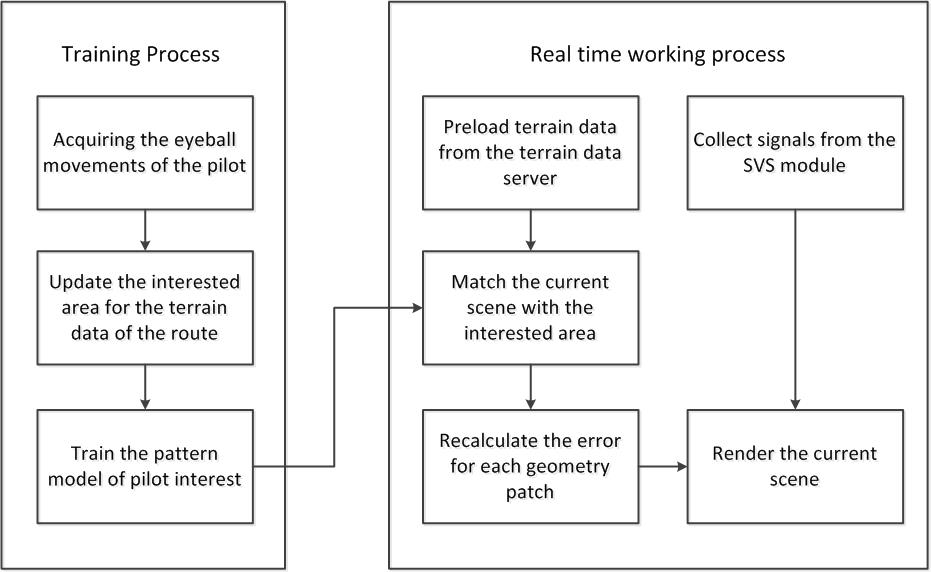}
\caption{Procedure of the SVS and Eyetracking modules}\centering
  \label{fig:e3dg}
\end{figure}




\subsection{Terrain Data Structure}

We propose a tree structure for the terrain data in order to integrate various types of geometry meshes into one hierarchical model. The improved terrain data structure is capable of emphasizing the interested area of the pilot.


First, we assume $P=\{p_1, p_2, ..., p_m\}$  be the data point sets within the $xy$-plane. Suppose for each data set $P$, the bounding box of $P$ is within the area surrounded by the extremal points $a_{min}$ and $a_{max}$. The domain of $P$ is defined as $D(P) \: = \lbrace (x,y) |x_{a_{min}} \leq x \leq x_{a_{max}}, y_{a_{min}} \leq y \leq y_{a_{max}}\rbrace$.

Let $Q := \{q_{u,v} | 1 \leq u \leq m; 1 \leq v \leq n\}$ be a texture mesh which consists terrain patches $q_{u,v}$ of $m \times n$. The extremal points of the axis parallel bounding box are denoted by $b_{min}$ and $b_{max}$. The domain $D(Q)$ of the texture $Q$ is defined by $D(Q) := \{(x, y)|x_{b_{min}} \leq x \leq x_{b_{max}}; y_{b_{min}} \leq y \leq y_{b_{max}}\}$.

We define a geometry patch in a tree structure as $T_{s,n}(P)$ for a terrain data set $P$, where there are at most $s$ points of terrain data within the boundary, and $n$ child nodes in the tree structure. Each child node $N$ represents a rectangular region $D_N(P_N) \subset D(P)$. The bounding box of $P_N$ is the domain $D_N$. The way of calculating the exact number of points for each node depends on the data preload strategy.



For every node $N$, there are at most $n_d$ child nodes contained. We construct the child nodes as follows: Let $D_N$ be the terrain data of a node N of which the domain $D_N \subset D(T)$. The geometric approximation error $e_D(N)$ are defined as the maximal vertical distance between the terrains D and $D_N$, i.e. \cite{1}, $e_T(N):= max_{p\in D_N} | h_D(p) - h_{D_N}(p)|$. For each subdomain $D_{N_i}$, a node $N_i$ will be constructed which approximates the terrain in that subdomain, and will be added as child node to the parent node N. If the error between $T$ and $T_N$ exceeds a certain threshold $\alpha \geq 0$, i.e., $e_T(N)> \alpha$, then the domain of $N$ is subdivided into a set of at most $d$ rectangular subdomains such that $D_N = \cup D_{N_i}$ with $D_{N_i} \cap D_{N_j} = \oslash, i \neq j, 1 \leq i, j \leq d$. The domain of the root node of the tree is $D(T)$ covering the whole domain of the terrain T.

\subsection{Data Preloading Algorithm}

In this section, we introduce the data preload algorithm that is utilized in the proposed integrated SVS system. Although the modern graphics stations may be capable of rendering thousands of shaded or textured polygons at interactive rates, the geometry complexity of the terrain data is still far more exceeded the capability of computation servers on the aircraft. Therefore, many algorithms of Level of Detailed (LOD) rendering have been proposed in the previous work of large terrain display on airbone SVS systems. The bottleneck of ordinary SVS systems is that the terrain data are too large to be stored in the computation servers. Only the data that may be required in the future will be loaded from the terrain data server to the computation server. In order to minimize the unnecessary transmitted data, we propose the data preload algorithm that is able to select the least required data to be preloaded.


The data preload algorithm considers the navigation information, the aircraft status, the operation of the pilot, and the important spots by tracking the eyeball movement of the pilot. In ordinary SVS systems, the terrain data are stored in a format of hierarchical bintree or quadtree. The vertexes of each triangle/rectangle are arranged in layers. Generally, when the SVS module needs to render the current view of the terrain, it renders the terrain near the flight in high resolutions, and the distant view in low resolutions. Since the capability of the computation server and the terrain data server is limited, at each time step, the computation server only loads the necessary terrain data from the terrain data server.

\begin{figure}[htb]
 \centering
  \includegraphics[width=0.6\columnwidth]{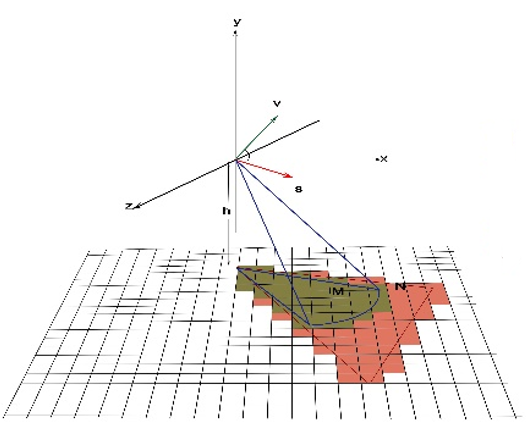}
\caption{Preloaded data according to the fusum}\centering
  \label{fig:e3dg}
\end{figure}

The navigation information of the aircraft is obtained from the navigation server of the SVS module. The navigation information contains the route from the source to the destination. The computation server only preload the terrain data in the sight scope of the route.

The flight status of the aircraft mainly consists of three kind of information: the altitude, the vocality, and the direction of the flight. In Figure 4, the area of preloaded terrain data of current time step is illustrated. We suppose that left angle and right angle to the forwarding direction is $\delta_l$ and $\delta_r$. $d$ is the extended terrain area of sight. In order to simplify the computation, we suppose the aircraft is the origin of axis, the aircraft heading is the Y axis, the vector connected the center of the earth and the aircraft is the Z axis, and the vector perpendicular to the YOZ plane is the X axis. The height of the aircraft is $h$. The velocity of the aircraft is $v$. At the initiation, the preloaded terrain area $d_0=\delta(h)$. $\delta_l$ and $\delta_r$ are defaulted angle that represent the possible area of sight. $\delta_l=\delta_r=\delta_0$. The preloaded terrain area is $S_0=d^2tan\delta_0$. The real time extended area of sight $d$ and the angles $\delta_l$ and $\delta_r$ are computed in formulation 1.

\begin{equation}\label{1}
\begin{cases}
d=\theta(h+v\times \delta t \times cos\theta)\\
\sigma_l=\sigma_0+\eta(v' sin\alpha)\\
\sigma_r=\sigma_0-\eta(v' sin\alpha)

\end{cases}
\end{equation}
%

In formulation 1, $\theta$ is the angle between the velocity vector and the Z axis. The real-time preloaded terrain area is $S=d^2(tan\delta_l+tan\delta_r)/2$. For each frame in the time step, $T_{S_v}$ is defined as the geometry patch of this terrain area $S$ at time $v$. Based on the tree structure defined previously, we preload the nodes of $T_{S_v}$ for each frame.

%

\subsection{Marking Interested Areas based on Eye Tracking}

The greedy data preload algorithm is capable of minimize the amount of unnecessary terrain data transmission between the terrain data server and the computation server. However, based on the survey of pilots encountering the abnormal events during the flight [15], each pilot requires different spots on the screen for higher resolution. Since there is no obvious way of satisfying all pilots' needs, we propose a method that discovering the important spots via the eyeball movements of pilot.

The novel method consists of two steps. First, the flight route and the eyeball movements of the pilot are recorded during the whole trip with time stamps. Each eyeball movement is re-mapped to the terrain data of the route by the time stamps.



According to the study of [16], the interested area can be acquired by learning the trial of eyeball movements. Since we use the time stamp to re-map the eyeball movements and the terrain vertexes, it is easy to find out the spots that the pilots gazed or become watchful in the route. It is also possible to use HMM model or other machine learning method to find out the interested area in a more precise way.

After the interested spots are located, the terrain data server preload these data to the computation server in the next flight trip. Each time, the interested area are alternated by the last record of the same trip. The mechanism may maintain a limited list of spots descendent by priority. The spots of least priority are removed from the list, and new spots of higher priority are added into the list. The whole algorithm is present in Algorithm 4.1.


\begin{algorithm}[!htbbp]
{\footnotesize \caption{The Data Preload Algorithm} \label{alg:RDPM}
\begin{algorithmic}[1]

\REQUIRE{ The terrain data points of trip $P$, the record of eyeball movement during the trip $E_P$, the priority list of previous time step $t-1$ $L_{t-1}$, the time stamps of the terrain data in the route $S_r$, the time stamp of the eyeball movement during the trip $S_e$ }

\ENSURE{ A new priority list of spots $L_t$. }
\STATE{Remap $E_t$ to $T$ based on $S_r$ and $S_e$.}
\STATE{Detect interested area of vertexes by the gazing time or the glance time. The interested spots are given a priority based on the attention time of the pilot.}
\STATE{Update the priority list $L_{t-1}$ by comparing the priority of new spots to the ones in the list. The spots of least priority are removed from the list.}
\STATE{Generate the new priority list $L_t$ by moving the top N spots of $L_{t-1}$ to the $L_t$.}
\end{algorithmic}
}
\end{algorithm}


%
%


\begin{figure}[htb]
 \centering
  \includegraphics[width=1.0\columnwidth]{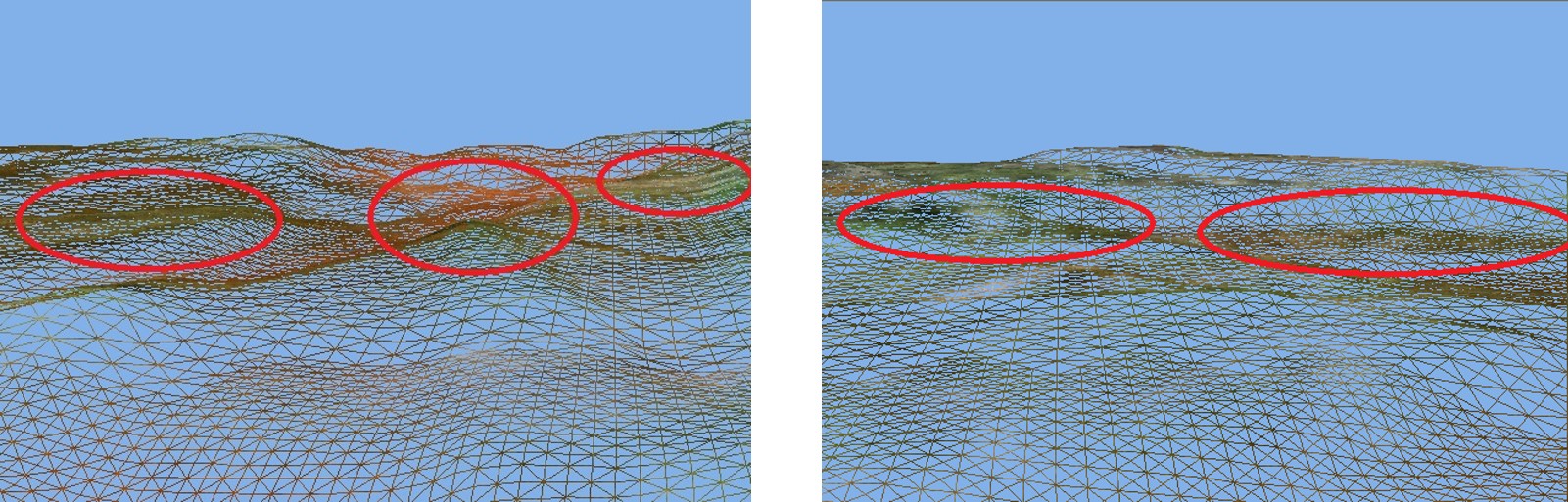}
\caption{The historical interested area (in red circles) of the pilots are displayed with higher resolutions.}\centering
  \label{fig:e3dg}
\end{figure}

\begin{figure}[htb]
 \centering
  \includegraphics[width=1.0\columnwidth]{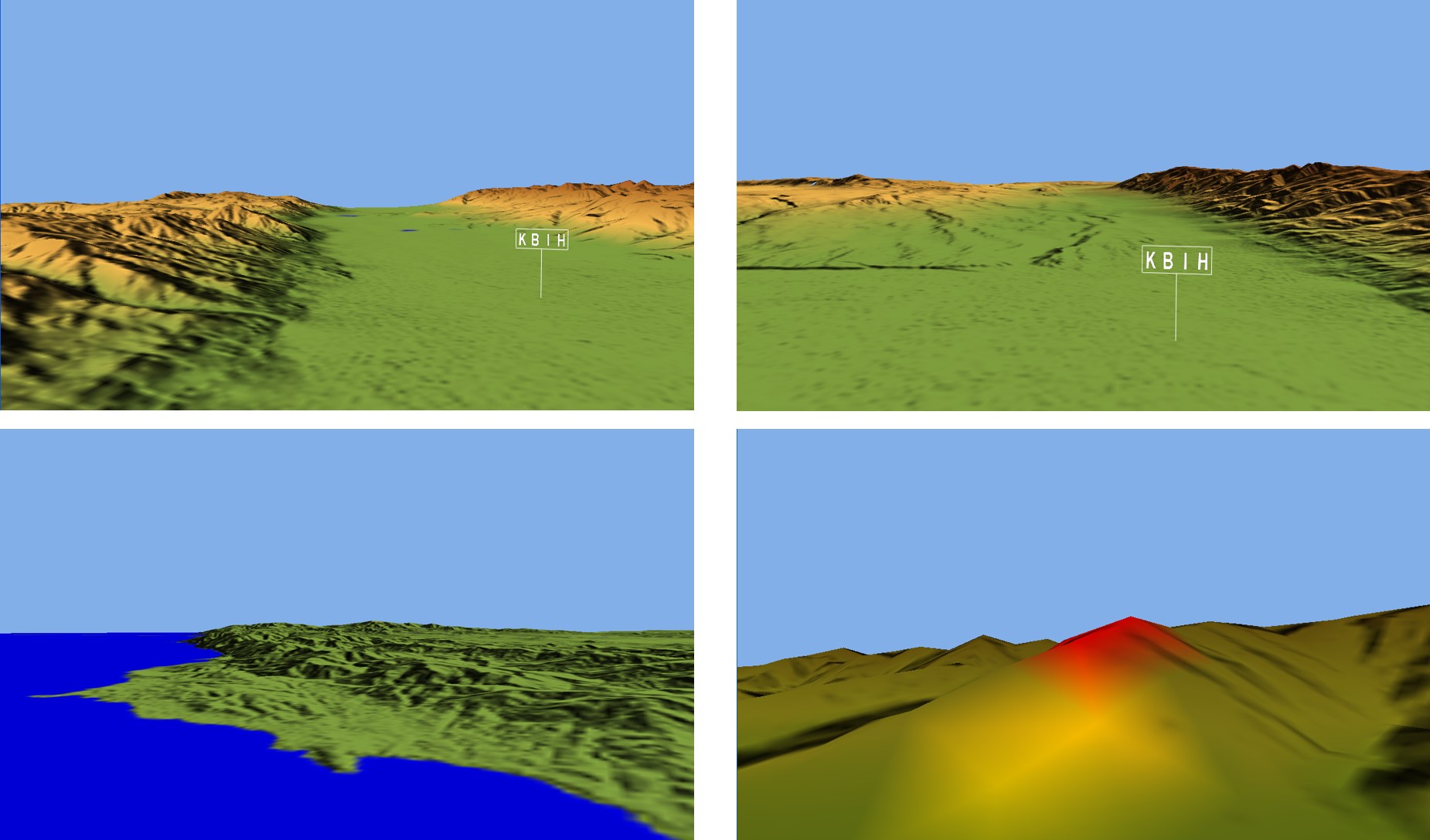}
\caption{The rendering result of the terrain area}\centering
  \label{fig:e3dg}
\end{figure}

\section{Fatigue Detection and Warning Mechanism}

In this section, we introduce the novel warning mechanism based on eye tracking. The warning mechanism, named as Eye Tracking Fatigue Alert (ETFA), is illustrated. The Proposed Algorithm can be divided into two steps. First, a fatigue detection mechanism is generated. Second, the fatigue generating mechanism is integrated with the SVS system.

In our implementation, every tenth frame from the eye tracking video is processed. We utilize the important spots that are acquired in the optimal data preload algorithm to activate the warning mechanism. When the flight is within the warning range of the spot, the fatigue alarm is activated. As long as the fatigue status of the pilot is above the normal status, an alert will be sent to the pilot.


When the system starts, frames are continuously fed from the camera to the eye tracking analyzer. We use the initial frame in order to localize the eye-positions. Once the eyes are localized, we start the tracking process by using information in previous frames in order to achieve localization in subsequent frames. During the tracking process, error detection is performed in order to recover from possible tracking failure. When a tracking failure is detected, the eyes are relocalized.

The fatigue detection and warning mechanism are activated during the flight simultaneously. When preloading the terrain data, the SVS system detects the potential risks on the terrain map, like high mountains, towers, waters, or other static obstacles. The sensors of the SVS system detects the flight status in real time in order to analyze the current risk level of the flight. If the pilots are in fatigue when the flight is near the risky terrain area, or the risk level of the flight is too high, the warning device will send signals immediately.

\section{Experiments and User Study}

The evaluations of our S-E system are conducted in three parts. First, we invited pilots using our S-E system in an aviating simulation environment in order to evaluate the display performance of the system. Second, we invited twenty pilots in 8 experimental scenarios to test the accuracy of fatigue detection. Third, we conduct a user study for the pilots for the evaluation of the improvement of driving experience. A questionnaire is given to the pilots who drive a long distance simulation using the proposed S-E system.

\subsection{Performance Evaluation}


The pilots's eyeball movement is recorded at the first time when they are participating the simulation. The multiresolution display performance is shown in figure 8.

The simulation is an one hour flight period, including the taking off period, the smooth flight period in the middle, and the landing period. We demonstrate the data transmission amount between the terrain data server and the computation server during the simulation in figure 9. In the table of the figure 9, the X axis is the time line and the Y axis is the amount of data transmitted from the data server to the computation server. The black line at the above is the amount of data transmitted without data prediction algorithm. The histogram is the amount of data transmitted utilizing the proposed data prediction algorithm. The ``Taking off'' title shows the time period of aircraft rising from the airport to a certain height. The ``Smooth flight in the middle'' title shows the time period that the aircraft cruises at certain altitude. The ``Landing'' title shows the time period of the aircraft descends to the ground.

\begin{figure}[htb]
 \centering
  \includegraphics[width=1.0\columnwidth]{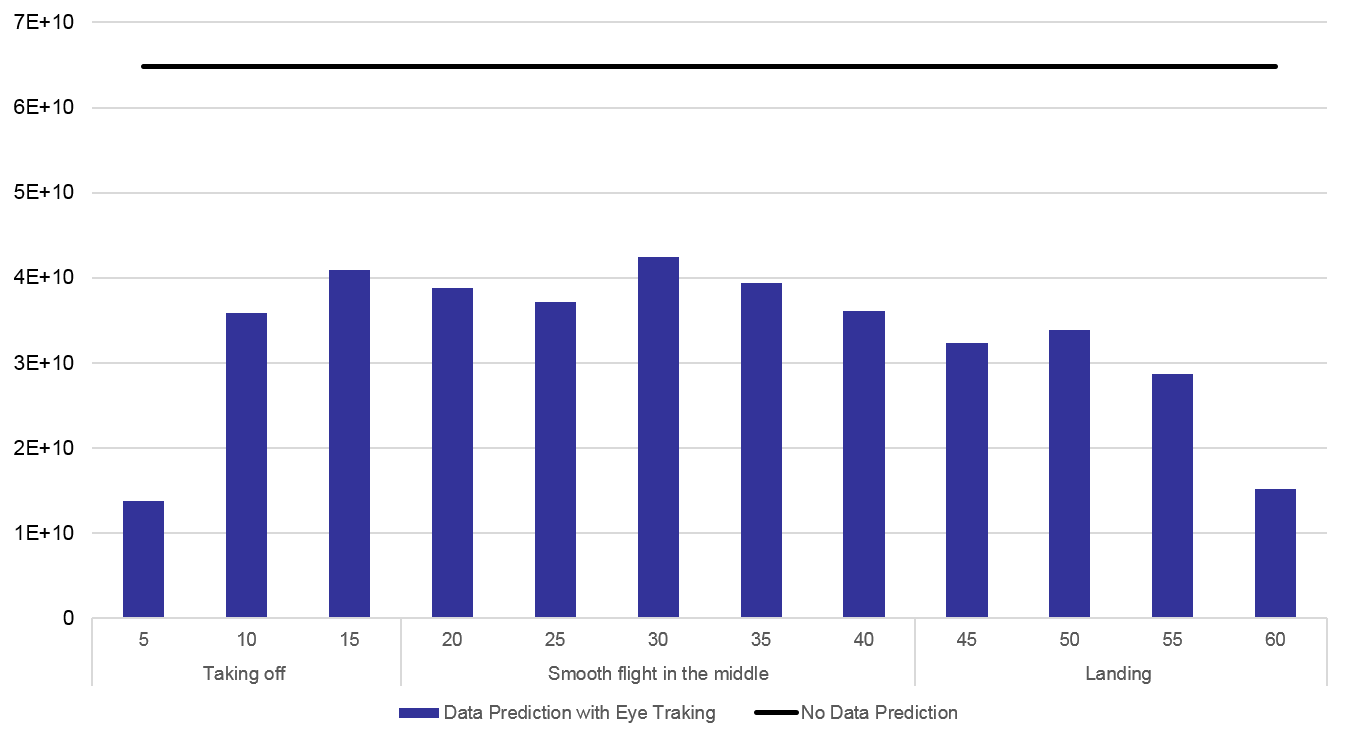}
\caption{Data throughput of the simulation }\centering
  \label{fig:e3dg}
\end{figure}

The evaluation shows that, without an optimized data preload algorithm, the throughput of the network and the computation load on the servers are remaining at a high status during the whole simulation. After utilizing the data prediction algorithm, the data transmission amount is reduced to an acceptable degree.

\subsection{Fatigue Detection Evaluation}


Twenty participants are invited for this evaluation. The experiment was conducted in a cockpit equipped with the proposed integrated SVS-Eyetracking system. 

Participants flew 8 experimental scenarios of 80-100 minutes each, involving a curved step-down approach through a terrain challenged region to a simulated airport.  Pilots were instructed to use the eye tracking device during the whole experiment period. While flying, pilots were instructed to detect any important/interest terrain spot that became visible on the SVS display, and also to report any changes to traffic altitude that they noticed on either the SVS display or the navigation display.

During the evaluation, some accidents are designed to occur at random time. If the pilot failed to react, the simulation will judge the pilot was at fatigue status. The eye tracking device also evaluates the fatigue degree of the pilot in real time. In all experimental scenarios, the eye tracking system did not warn the pilots when it finds out the pilots are in fatigue. The experimental results are shown in Table 1.

\begin{table}  \centering
  \begin{tabular}{|c|c|c|c|}
  \hline
      &No. of S-E&Total No.& Accuracy\\
  Scenario 1 & 55 & 97&56.7\%\\
  Scenario 2 & 68 & 90&75.6\%\\
  Scenario 3 & 34 & 37&91.9\%\\
  Scenario 4 & 60 & 75&80\%\\
  Scenario 5 & 23 & 24&95.8\%\\
  Scenario 6 & 17 & 17&100\%\\
  Scenario 7 & 25 & 28&89.3\%\\
  Scenario 8 & 9 & 10&90\%\\
\hline
  \end{tabular}
  \caption{Experimental results of 8 scenarios}
\end{table}

The first four scenarios are comparable challenging, and the possibilities of the occurring of risky scenarios are high. However, since the eyetracking device only record the fatigue warning signals, the number of fatigue detected by the S-E system is always smaller than the number of fatigue the simulation records. The experimental results of the last four scenarios shows that the number of fatigue that the S-E system detected are very similar to the number of fatigue the simulation detected.

\begin{table}[htbp]
\centering
\label{tab:parameter}
{\small
\begin{tabular}{|l|}
\hline
\textbf{Questions ratings from 1 to 5,} \\
\textbf{5 is the score for perfect experience}\\
\hline

Is the SVS system showing the terrain consistent with \\
the simulation environment?\\ \hline
Is the SVS system correctly showing the symbols of \\
particular spots on the map? \\ \hline
Is there any delay on the display screen of SVS system \\
during the simulation? \\ \hline
Is there any ambiguous symbols on the screen? \\ \hline
Is there any display error occurred during \\
the simulation? \\ \hline
How intrusive when you drive with the eye\\ tracking device?  \\ \hline
Is there any discomfort (e.g. dazzle) caused\\ by the system
during the simulation? \\ \hline
Is the warning signal noticeable? \\ \hline
Is there any potential risk are noticed by the alert? \\ \hline
Is the fatigue detection accurate according \\to the scenario? \\ \hline
Is the eye tracking device working correctly during \\
the simulation?\\ \hline
Is the interested area on the screen has better\\ resolution? \\ \hline
Is the interested area on the screen are the \\spots you have
been paying attention during\\ last simulation? \\ hline
Is there any fake alert during the simulation? \\ \hline
Is the terrain warned on the screen displays correctly? \\
Is the fatigue detection helpful for avoiding \\potential risks? \\ \hline


\hline
\end{tabular}}
\caption{Rating questions for driving experience}
\end{table}

\subsection{User Study for the Driving Experience}

The study in \cite{23} categorized the human factors issues related to SVS systems into three research areas: Image quality, information integration, and operational concepts. Based on their study, a questionnaire is given to the participants for the evaluation of their driving experience using the proposed S-E system.

The questionnaire consists up to thirty questions, which contains fifteen rating questions for the user's evaluation of their driving experience, and fifteen questions for their opinions of system setting and suggestions. The evaluation questions includes the evaluation of the performance of the SVS system, the comfort of using the system, the accuracy of fatigue detection and warning signal, and other question related to human factor. The questions of setting and opinions inquired the setting parameters of the system and suggestions of the participants. The questionnaire is shown in Table II and Table III.

Most of the pilots show that the system works smoothly with nearly no error on the display screen. Some participants indicates that the symbol of certain terrain spots are too simple, and expected us to update the textures of building and other obstacles. Some participants reported some fake alarms during the flight, but they also indicated that they didn't cause any maloperations. The overall feedback indicates the S-E system works efficiently and is helpful for avoiding the potential risks during the simulation.

The suggestions of the participants are carefully collected and the system setting is altered according to their opinions. The suggestions also indicates that the S-E system may require more consideration of human factors in the future.

\section{Conclusion}

In this paper, we propose an SVS-Eyetracking architecture which integrated the eye tracking device into an SVS system. Based on the proposed architecture, we proposed a data prediction algorithm which reduces the amount of data transmitted from the terrain data server to the computation server. The data prediction algorithm considers the flight status provided by the SVS system and the eyeball movement of the pilot. The evaluation shows that the proposed algorithm may reduce the data transmitted significantly. We also implemented an fatigue warning mechanism in the proposed S-E system. The evaluation shows that the S-E system works efficiently and is capable of avoiding potential risked caused by fatigue in the flight simulation.

\section*{Acknowledgment}

The authors would like to thank...

\ifCLASSOPTIONcaptionsoff
  \newpage
\fi

\end{document}